\documentclass[preprint,12pt,epsfig]{revtex4}

\usepackage {graphicx}

\usepackage {amsmath}

\usepackage {amsfonts}

\usepackage {amssymb}

\usepackage {mathrsfs}

\usepackage{color}

\newcommand {\be}{\begin{eqnarray}}

\newcommand {\ee}{\end{eqnarray}}

\begin{document}

\title{Analytical derivation of thermodynamic properties of bilayer membrane with interdigitation.}
\author {Sergei I.\ Mukhin}

\email {sergeimoscow@online.ru}

\affiliation {Theoretical Physics Department, Moscow Institute for Steel \& Alloys, Moscow, Russia}
	
\author {Boris B.\ Kheyfets}

\affiliation{Physical Chemistry Department, Moscow Institute for Steel \& Alloys, Moscow, Russia}

\date {\today}
%\draft

\begin{abstract}
We consider a model of bilayer lipid membrane with interdigitation, in which the lipid tails of the opposite monolayers interpenetrate. The interdigitation is modeled by linking tails of the hydrophobic chains in the opposite monolayers within bilayer as a first approximation. A number of thermodynamical characteristics are calculated analytically and compared with the ones of a regular membrane without interdigitation. Striking  difference between lateral pressure profiles at the layers interface for linked and regular bilayer models is found. In the linked case, the lateral pressure mid-plane peak disappears, while the free energy per chain increases. Within our model we found that in case of elongation of the chains inside a nucleus of e.g. liquid-condensed phase, homogeneous interdigitation would be more costly for the membrane's free energy than energy of the hydrophobic mismatch between the elongated chains and the liquid-expanded surrounding. Nonetheless, an inhomogeneous interdigitation along the nucleous boundary may occur inside a ``belt'' of a width that  varies approximately with the hydrophobic mismatch amplitude.
\end{abstract}
% New page

\maketitle

\section{Introduction}
\label{sec:Introduction}

Studying mechanisms of changes in the structure of cell membrane under the adsorption of small amphiphilic molecules (alcohol, anesthetics, etc.) is of fundamental interest, as well as is important for understanding of the functioning of the cell membranes and embedded proteins \cite{kranenburg}. One of the drastic changes of the structure is membrane transition into the interdigitated phase \cite{intosh}. While in a regular membrane the thickness of the hydrophobic part of a bilayer is approximately twice the length of the hydrophobic tails of the phospholipid, an interdigitation may reduce the hydrophobic thickness to the sum of the length of the lipid tail and of the small amphiphilic molecule. Understanding possible consequences of the interdigitation for the lipid membrane properties is important for e.g. prediction of the effects of anesthetics on the functioning of the ion channels embedded in the membrane \cite{Cantor},\cite{tang}.

\begin{figure}[h]
	\centering
		\includegraphics[height=2.5 in]{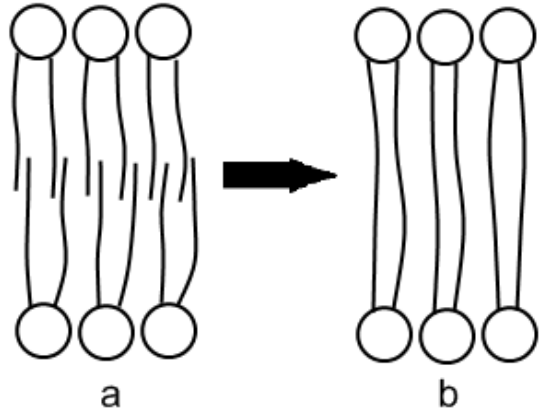}
	\caption{Model of membrane with interdigitation we based on.}
	\label{fig:int_model}
\end{figure}

As a first approximation to the lipid bilayer membrane with interdigitation, in which lipid tails from the opposite monolayers interpenetrate, we consider a model with pairwise linked tails of the lipids belonging to the opposite monolayers within a single bilayer, Fig.~\ref{fig:int_model}. Our model does not allow for a lateral area dilation of the membrane, that may follow the interdigitation \cite{kranenburg}, but it bears an important property of the interdigitated bilayer in the form of constrained meandering freedom of the chains ends in the vicinity of the monolayers interface. We found important consequences of this constriction: the entropy of the bilayer decreases, the free energy increases, and the lateral pressure profile $\Pi_t(z)$ changes drastically.

\begin{figure}[h]
	\centering
		\includegraphics{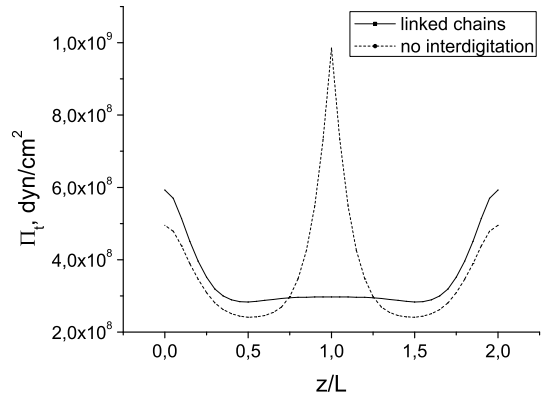}
	\caption{Lateral pressure distribution in the hydrophobic core of the bilayer. $z$ is coordinate along the chain axis normalized by the monolayer thickness $L$ and spanning from one head group ($z=0$) to another ($z=2L$). The parameters for the lipid bilayer are as follows: monolayer thickness $L=15 A$, area per chain $A_0=20 A$, chain flexural rigigity $K \sim kTL/3$, temperature $T=300 K$, total monolayer pressure $P=70$ $dyn/cm$.}
	\label{fig:presprof12}
\end{figure}

\begin{figure}[h]
	\centering
		\includegraphics{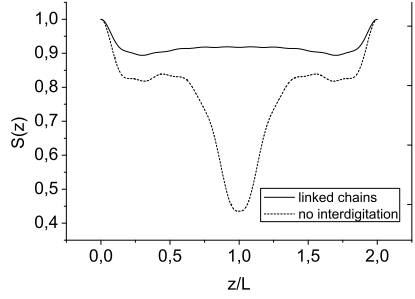}
	\caption{Order parameter in cases of linked chains (see Fig.~\ref{fig:int_model}) and no interdigitaion (see Fig.~\ref{fig:effect}).  }
	\label{fig:sz_12}
\end{figure}

The two distinct $\Pi_t(z)$ curves, see Fig.~\ref{fig:presprof12}, can be understood by comparing orientational fluctuations of the hydrocarbon segments of the semi-flexible lipid chains in the interdigitated- and noninterdigitated lipid bilayers. These fluctuations can be characterized by an orientational order parameter $S(z)$, see Fig.~{\ref{fig:sz_12}} ($z$ is coordinate measuring depth inside bilayer), calculated using our model. 
The fluctuations reach their maximum at the monolayers interface inside a noninterdigitated bilayer, because the chains ends are free there. Hence, the order parameter drops at $z=L$, see the dashed curve in Fig.~{\ref{fig:sz_12}}. Simultaneously, a maximum of the entropic lateral pressure occurs at $z=L$, dashed curve in Fig.~\ref{fig:presprof12}. Distinctly, in the interdigitated bilayer fluctuations are significantly suppressed at the monolayers interface due to restriction of orientational freedom of the central segments by their peripheral neighbors. Hence, $S(z)$ does not drop at $z=L$, see solid curve in Fig.~{\ref{fig:sz_12}}. As a consequence, there is no maximum at $z=L$ in the $\Pi_t(z)$ dependence drawn with solid line in Fig.~\ref{fig:presprof12} for interdigitated bilayer case.

We present our analytical results derived in closed form for thermodynamical properties of a membrane with linked chains in the weak interdigitation limit: i.e. thickness of the hydrophobic part of a bilayer is comparable with twice the length of a hydrophobic chain. In this limit we use more complete version of the energy functional entering the membrane partition function than developed earlier \cite{mubao}: besides the bending energy of a chain conformation, we included kinetic energy  of the lipid chain. We prove that this makes the path integral representation of the free energy of the chains uniquely normalizable.

The plan of the article is as follows. In Sec. II we introduce a microscopic model of a membrane with interdigitation and calculate the membrane free energy using path-integral summation over the chains conformations. The inter-chain entropic interactions are treated in the mean-field approximation. Several thermodynamic moduli characterizing the interdigitated bilayer are derived as well and compared with noninterdigitated case. An increment of the free energy per chain due to interdigitation is calculated. In Sec. III we calculate analytically the lateral pressure distribution (profile) across the hydrophobic core of the lipid bilayer and make comparison between the cases with- and without interdigitation. Also calculated is chain order parameter that characterizes correlations between the orientations of the chain segments and clearly demonstrates an increase of the orientation order in the interdigitated case as compared with noninterdigitated bilayer.  In the Appendix we discuss possibility of inhomogeneous interdigitation along the boundary of the transmembrane liquid-condensed domain (raft) embedded in the liquid expanded surrounding in the bilayer membrane. We evaluate energetically favorable configuration of such a raft allowing for the trade between hydrophobic mismatch and interdigitation-induced free energy increase.

\section{Microscopic model of interdigitated bilayer}

\begin{figure}[h]
	\centering
		\includegraphics{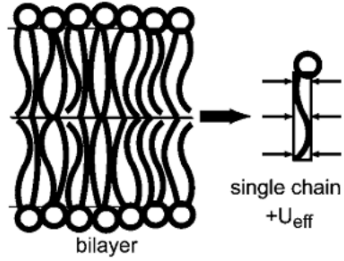}
	\caption{Model of lipid membrane in the mean-field approximation: we substitude interaction between neightboring chains by an effective quadratic potential.}
	\label{fig:effect}
\end{figure}

The interdigitated lipid bilayer membrane is modeled by linking pairwise the tails of the chains belonging to the opposite monolayers. Hence, a couple of linked chains is substituted by a single semi-flexible string of length $\approx 2L$,
where $L$ is the monolayer thickness, see Fig.~\ref{fig:effect}. Correspondingly, conformations of the string as a ``trans-membrane'' object obey combined boundary conditions at the opposite head group regions of bilayer with coordinates $z=0$ and $z=2L$ respectively, as is described below in detail.

With bending (flexural) rigidity $K_f$, and with the mean-field approximation accounting for entropic repulsion between neighboring couples of pairwise linked chains (see Fig.~\ref{fig:effect}), the energy functional of a single string, $E_t$, has the form:

\begin{equation} \label{eqn:et}
   E_t=\int ^{2L}_{0} \left\{ \frac{\rho \dot{\textbf{R}}^2(z)}{2}+\frac{K_f}{2} \left( \frac{\partial^2 \textbf{R} (z)}{\partial z^2} \right )^2 +         \frac{B}{2}\textbf{R}^2(z) \right\} dz.
\end{equation}

\noindent 
Here harmonic potential $U_{eff}=BR^2/2$, with self-consistently defined rigidity $B$, describes entropic repulsion between the strings, $z$ is coordinate along the string axis, and $\textbf{R}(z)$ is vector in the $\left\{x,y\right\}$ plane characterizing deviation of the string from the straight line, 
$\textbf{R} ^2 = R_x^2+R_y ^2$. The choice of harmonic potential is justified since we assume finite "softness" of the effective ``cage'' created by the neighboring lipid chains in the limit of small chain deviations. A harmonic potential was considered in earlier work \cite{Burk} for a semi-flexible polymer confined along its axis. The first term in Eq. (\ref{eqn:et}) represents kinetic energy of the string, $\rho$ is linear density of mass: $\rho=m(CH_2)N/L$, where $m(CH_2)$ is a hydrocarbon group mass, N is the number of hydrocarbon groups per chain (for numerical estimates we took $N=9$, see \cite{Rubin}).

The bending energy term in Eq. (\ref{eqn:et}) represents the energy of the chain trans or gauche conformations. It contains the second derivative over the $z$ coordinate rather than over the contour length of the chain. This approximation is valid provided that deviations from the $z$ axis are small with respect to the chain length $L$:  

\begin{equation} \label{eqn:2}
   \frac{\sqrt{\left\langle \textbf{R}^2(z)\right\rangle}}{2L} \le \left(\frac{k_B T}{L^2 P_{eff}} \right)^{1/2} \ll 1.
\end{equation}

\noindent
This limit is opposite to the one considered in the long polymer theory \cite{Nelson}, where the second derivative in Eq.~(\ref{eqn:et}) is substituted by the first derivative in the flexible chain approximation.

Using the functional Eq.~(\ref{eqn:et}) the chain partition function is found as a path integral over all string conformations:

\begin{equation} \label{eqn:4}
   \begin{split}
   Z=\int{exp\left(-\frac{E \left((\dot{\textbf{R}}(z),\textbf{R}(z))\right)}{k_B T}\right)D\dot{R}_x DR_x D\dot{R}_y DR_y}=
   \\
   \left(\int exp\left[-\frac{E\left((\dot{R}_x(z),R_x(z))\right)}{k_B T}\right]D\dot{R}_x DR_x\right)^2=Z_x^2
\end{split}
\end{equation}

\noindent
The second equality in Eq.~(\ref{eqn:4}) holds when the membrane is laterally isotropic and the x and y deviations can be considered independently.

To calculate the path integral Eq.~(\ref{eqn:4}) we rewrite the energy functional Eq.~(\ref{eqn:et}) using the self-adjoint operator $\hat{H}$:

\begin{equation} \label{eqn:5}
   E_{t}= \sum _{i=x,y} \frac{1}{2}\int^{2L}_{0} {  \left( \rho \dot{R_i}^2(z)+R_i(z) \hat{H} R_i(z)dz\right)},
\end{equation}

\begin{equation} \label{eqn:6}
   \hat{H} = K_f \frac{\partial^4}{\partial z^4} + B.
\end{equation}

The operator $\hat{H}$ is obtained after integrating by parts the expression Eq.~(\ref{eqn:et}) under the following boundary conditions for the string that models two linked chains belonging to the opposite monolayers (the z-coordinate spans from one head group at $z=0$ to another at $z=2L$).
The chain angle is fixed in the head group region:
\begin{equation}
\label{eqn:7a}
%\begin{split}
R'(0)=0;\, 
R'(2L)=0
%\end{split}
\end{equation}
\noindent
No total force is applied upon chain at the head group:
\begin{equation}
\label{eqn:7b}
%\begin{split}
R'''(0)=0; \,  
R'''(2L)=0
%\end{split}
\end{equation}

\noindent
These boundary conditions, as well as the energy functional in Eq.~(\ref{eqn:et})differ from the ones used to describe a single monolayer of a noninterdigitated lipid bilayer (compare \cite{mubao}):

\begin{equation} \label{eqn:etn}
   E^m_t=\int ^{L}_{0} \left\{ \frac{\rho \dot{\textbf{R}}^2(z)}{2}+\frac{K_f}{2} \left( \frac{\partial^2 \textbf{R} (z)}{\partial z^2} \right )^2 +         \frac{B}{2}\textbf{R}^2(z) \right\} dz
\end{equation}

\noindent
where $E^m_t$ is the energy functional of a single monolayer, and the motions of the chains in the opposite monolayers forming a noniterdigitated bilayer are independent.  The total energy of a bilayer in this approximation is then twice the energy of a single monolayer: $2\times E^m_t$. We impose the following boundary conditions for a monolayer:
the chain angle is fixed, and no total force is applied 
to the chain at the head group:
\begin{equation}
\label{eqn:7ma}
%\begin{split}
R'(0)=0;\, 
R'''(0)=0
%\end{split}
\end{equation}
\noindent
No total force and no torque is applied at the free chain end (i.e. at the monolayers interface inside the bilayer):
\begin{equation}
\label{eqn:7mb}
%\begin{split}
R'''(L)=0;\,  
R''(L)=0
%\end{split}
\end{equation}

\noindent

Finally, in both cases, the free energy of a bilayer equals $F=-k_B T\ln(Z)$, where $Z$ is partition function of a bilayer. 
Using expressions Eq.~(\ref{eqn:et}) or Eq.~(\ref{eqn:etn}) for interdigitated or noninterdigitated bilayer respectively, we differentiate the free energy and obtain the self-consistency equation in the form (showed for interdigitation case):

\begin{equation}
	\label{eqn:13}
	\frac{\partial F}{\partial B}=2L\left\langle R^2\right\rangle
\end{equation}

\begin{figure}[h]
	\centering
		\includegraphics{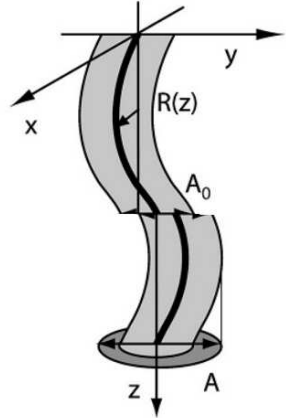}
	\caption{Hydrocarbon chain as a flexible string of finite thickness. $\textbf{R}(z)$ is the vector characterizing the deviation of the center of the chain cross section from the $z$ axis, $|\textbf{R}(z)| = \sqrt{R_x(z)+R_y(z)}$; $A_0$ is the "incompressible area" of the chain cross section; $A=\pi \left\langle \textbf{R}^2\right\rangle$ is the area swept by the centers of chain cross sections; $A$ is the average area per lipid chain in the bilayer.}
	\label{fig:nano}
\end{figure}

\noindent
As before \cite{mubao}, we take into account that hydrocarbon chains of lipid molecules are bulky objects that possess finite thickness and introduce an "incompressible area" of the chain cross section $A_0$ (see Fig.~\ref{fig:nano}). The area occupied by a lipid chain in the bilayer is related to the string mean square deviation $\left\langle \textbf{R}^2\right\rangle$  by the following formula \cite{mubao}:

\begin{equation} \label{eqn:a}
   \delta A = \pi \left\langle \textbf{R}^2\right\rangle = \left(\sqrt{A} - \sqrt{A_0}\right)^2,
\end{equation}
\noindent
where $\delta A$ is the area swept by the string formed with the centers of the chain cross sections. In the text below we imply by chain deviations those of a string described by the $\textbf{R}$ vector.
The self-consistency equation Eq.~(\ref{eqn:13}) combined with formuli Eq.~(\ref{eqn:a}) permits us to find the $A$ dependence of the coefficient of entropic repulsion $B$ and finally derive the membrane equation of state in a form of pressure-area isotherm (see Appendix for details).

To make numerical estimates based on our model of a lipid bilayer we use the following parameters values: chain length $L=15 A$, chain incompressible area $A_0 =20 A^2$, $T_0 =300 K$ as reference temperature. The chain flexural rigidity is defined as
\cite{landau} $K_f=EI$, where $E\approx 0.6 GPa$ is the chain Young's modulus \cite{polyna} and $I=A_0^2/4\pi$ is the (geometric) moment of inertia. The flexural rigidity can also be evaluated from polymer theory \cite{Nelson} $K_f=kBTl_p$, where $l_p\approx L/3$ is the chain persistence length \cite{polyna} and $k_B$ is the Boltzmann constant. Both estimates give approximately $K_f \approx k TL/3$ at chosen $L$ and at $T=T_0$.

\section{Interdigitated bilayer: the free energy increment}
The eigenvalues and eigenfunctions of the operator $\hat{H}$ defined in Eq.~(\ref{eqn:6}) obey the following equation:

\begin{equation}
	\label{eigens}
%\begin{split}
	\hat{H}R_n\equiv K_f \frac{\partial^4R_n}{\partial z^4} + BR_n=E_nR_n
%\end{split}	
\end{equation}

\begin{figure}[h]
	\centering
		\includegraphics{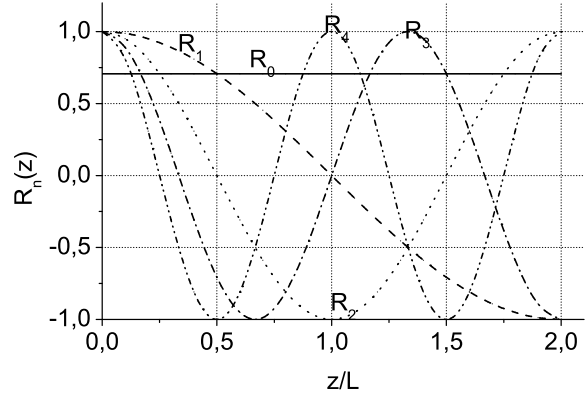}
	\caption{The eigenfunctions $R(z)$ of the self-adjoint operator $\hat{H}$ for the boundary conditions  Eq.~(\ref{eqn:7a}) and Eq.~(\ref{eqn:7b}). Other parametrs are as in Fig.~\ref{fig:presprof12}.}
	\label{fig:eigens_int}
\end{figure}

 \noindent
 Solving this equation with the boundary conditions Eq.~(\ref{eqn:7a})-(\ref{eqn:7b}) one obtains:

\begin{equation}
	\label{eqn:8}
%\begin{split}
	E_n=B+\frac{k_n^4 K_f}{L^4}, k_n=\pi n/2,  n\ge 1;\; E_0=B,
%\end{split}	
\end{equation}

\begin{equation}
	\label{eqn:9}
	%\begin{split}
	R_n(z)=c_n \cos (k_n z/L), n\ge 1; \; R_0(z)=\sqrt{\frac{1}{2L}},
	%\end{split}
\end{equation}

\noindent
where $c_n=\sqrt{1/L}$ and $\lambda_n=2\pi L/k_n$ is the wavelength. Several eigenfunctions are shown in Fig.~\ref{fig:eigens_int}.

Then an arbitrary string conformation, described with the deviation from the $z$-axis, $R_x(z,t)$, as well as its  energy are expanded over eigenfunctions $R_n$ and eigenvalues $E_n$ found from Eq.~(\ref{eigens}):

\begin{equation}
	\label{eqn:10}
	\begin{split}
	R_x(z,t)=\sum_{n=0}{C_n(t) R_n(z)};\\
	\dot{R}_x=\sum_{n=0}{\dot{C}_n R_n}; \;\;\;
	E_t=\frac{1}{2}\sum_{n=0}{\rho \dot{C}_n^2 + C_n^2 E_n}
	\end{split}
\end{equation}

The bilayer partition function is then found as the integral over the coefficients of expansion $C_n$ and conjugated momenta $p_n=\rho\dot{C}_n$ in Eq.~(\ref{eqn:10})

\begin{equation}
\label{eqn:11}
	\begin{split}
Z_x=\int^{\infty}_{-\infty} {\prod_{n=0} {\exp\left(- \frac{p_n^2}{2\rho k_B T}-\frac{C_n^2E_n}{2 k_B T} \right)\frac{dp_n dC_n}{2\pi \hbar}}}=
\\
=\prod_{n=0} \frac{k_B T}{\hbar} {\sqrt{\frac{\rho }{E_n}}}=\prod_{n=0}\frac{k_B T}{\hbar \omega_n}
\end{split}
\end{equation}

\noindent
where $\omega_n=\sqrt{E_n/\rho}$. It is important that the latter expression for $\omega_n$ in the limit of a free string, $B\equiv 0$, leads to the well known bending waves spectrum of Euler beam \cite{landau}: $\omega_n=\sqrt{EI\tilde{k}_n^4/\rho}$ (with $\tilde{k}_n\equiv k_n/L$), as it follows from Eq.~(\ref{eqn:8}) and expression for the bending rigidity $K_f=EI$ mentioned above. Hence, by including kinetic energy of the chain into the energy functional $E_t$ we obtain correct dimensionless expression for partition sum in the Eq.~(\ref{eqn:11}). 

Using  Eq.~(\ref{eqn:11}), Eq.~(\ref{eqn:4}) and $F=-k_B T \ln {Z}$ we find the following expression for the free energy of a bilayer with interdigitation in our model: 
\begin{equation}
	\label{eqn:26}
	F_{int}=- 2k_B T\sum_{n=0}^{n_{max}}\ln {\frac{k_B T}{\hbar \omega_n}}
\end{equation} 
 \noindent
 Using then relation $S_{int}=-\partial F_{int}/\partial T$ we find the following expression for the entropy $S_{int}$:

\begin{equation}
	\label{eqn:27}
	S_{int}=- \left(\frac{\partial F}{\partial T}\right)_V=-2k_B \sum_{n=0}^{n_{max}} \left[{\ln \left(\frac{k_B T}{\hbar \omega_n}\right)}   + T\left\{    \frac{1}{\omega_n} \left(\frac{\partial \omega_n}{\partial T}\right)_V - \frac{1}{T}\right\}\right]
\end{equation} 

\noindent 

Both expressions are valid provided the motion of the lipid chains at room temperature $T$ is classical (not quantum), i.e. : $k_B T / \hbar \omega_{n_{max}}>>1$ and upper cutoff $n_{max}$ in the sums is defined by condition that the shortest half-wavelength $0.5\lambda_{n_{max}}=\pi L/k_{n_{max}}$ of the eigenfunction $R_{n_{max}}$ is not shorter than the $CH_2$-monomer length of the ``chain segment''.

\begin{figure}[h]
	\centering
		\includegraphics{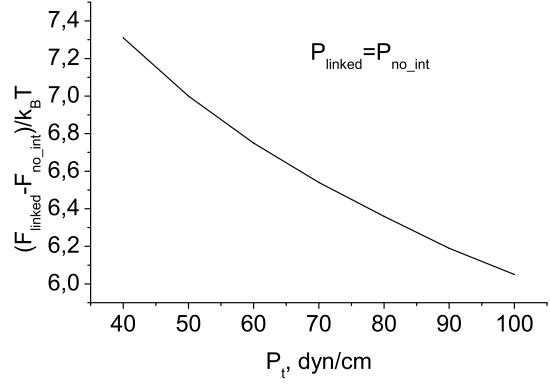}
	\caption{Not normalized free energy (per chain) difference  of membranes with linked chains and with no interdigitation. Linked chains membrane ``cost'' more free energy.}
	\label{fig:F1}
\end{figure}

\begin{figure}[h]
	\centering
		\includegraphics{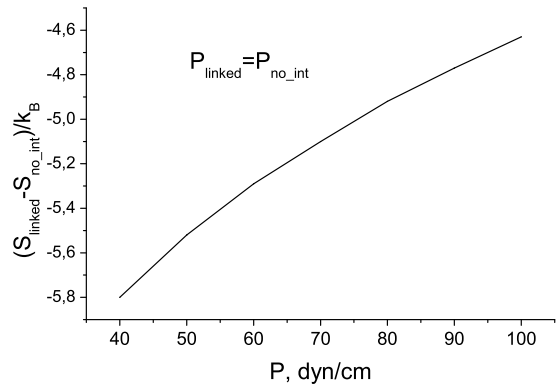}
	\caption{	Entropy (per chain) difference between membranes with linked chains and with no interdigitation. Entropy of linked chains membrane is lower that of no interdigitation membrane.}
	\label{fig:S1}
\end{figure}

The free energy and entropy of the noninterdigitated bilayer, $F_{non}$ and $S_{non}$ respectively, are obtained using the same relations as in Eqs.~(\ref{eqn:26}) and (\ref{eqn:27}), but with the corresponding change of the frequencies spectrum $\omega_n$ that results from the noniterdigitated bilayer conditions expressed in Eqs.~(\ref{eqn:7ma})-(\ref{eqn:7mb}). Using the above relations we calculated interdigitation free energy and entropy ''cost'' as the differences of the respective bilayer free energies and entropies in the interdigitated and noninterdigitated cases. Our results are represented in Fig.~\ref{fig:F1} and Fig.~\ref{fig:S1}. In Fig.~\ref{fig:F1} the free energy increment (per chain) of the order of $5k_BT$ in the intedigitated bilayer with respect to the non-interdigitated one is caused by the corresponding decrease of the entropy $\sim 5k_B$ (per chain), see Fig.~\ref{fig:S1}. Location of the entropy decrease in the interdigitated bilayer can be found by exploring the chain's orientational order parameter $S(z)$ defined as:

\begin{equation}
	\label{eqn:S}
	S(z)=\frac{1}{2}(3\langle cos^2\theta (z)\rangle -1)\,,
\end{equation} 
\noindent where $\theta(z)$ gives distribution of the tangent angle of the chain across the bilayer. Straight (ordered) chain possesses $\theta\equiv 0$ and $S(z)\equiv 1$. In the limit of small deviations from the straight line $\theta \leq 1$ considered in our model the order parameter can be expressed using the following relations:
\begin{equation}
	\label{eqn:tg}
	\langle cos^2\theta (z)\rangle\approx 1-\langle tg^2\theta (z)\rangle=\langle (R' (z))^2\rangle=\frac{k_BT}{2}\sum_{n=0}
	\frac{(R_n'(z))^2}{E_n}\,,
\end{equation} 
so that finally we obtain:
\begin{equation}
	\label{eqn:S}
	S(z)\approx 1-\frac{3k_B T}{4}\sum_{n=0}\frac{(R_n'(z))^2}{E_n}\,.
\end{equation}

\noindent Calculated order parameter distributions across the bilayer, $S(z)$,in our model with and without interdigitation are represented in Fig.~{\ref{fig:sz_12}}. The solid line corresponds to linked chains (modeling an interdigitation), and dashed line is calculated for non-interdigitated case. It is obvious from the Fig.~{\ref{fig:sz_12}} that main difference occurs at the monolayers interface ($z=L$) inside the bilayer. Free chain ends acquire maximal disorder in this region, while linked tails remain quite ordered. Another manifestation of this mid-bilayer ordering phenomenon will be seen in the next section in the calculated behavior of the lateral pressure profile inside bilayer.

\section{Lateral pressure profile and pressure-area isotherms for interdigitated bilayer}

The equation of state of the lipid chains in the bilayer can be derived as follows:
\begin{equation}
\label{eqn:14}
    P_t=-\left(\frac{\partial F_t}{\partial A}\right)_T,
\end{equation}

\noindent
where $P_t$ is the total lateral pressure (or tension), produced by linked hydrocarbon chains. Substituting expression for the free energy from Eq.~(\ref{eqn:26}) into Eq.~(\ref{eqn:14}) one finds:

\begin{equation}
\label{eqn:lz}
    P_t=-k_BT\sum_{n=0}\frac{\partial E_n}{\partial A}\frac{1}{E_n}.
\end{equation} 

\noindent
We may consider $P_t$ as an integral of the lateral pressure distribution (profile) function, $\Pi_t(z)$, over the hydrophobic thickness of the bilayer:

\begin{equation}
\label{eqn:lppdef}
    P_t\equiv \int\Pi_t(z)dz.
\end{equation} 

\noindent
In order to find out $\Pi_t(z)$ defined this way, it is possible to use the following formal trick. 
Namely, the dependence on aria $A$ of $E_n$ arises via dependence of the ``potential'' $B(A)$, that enters operator $\hat{H}$ in Eq.~(\ref{eigens}). One may in addition formally consider $B(A)$ as being $z$-dependent function. Then, a well known relation from the perturbation theory \cite{LL3} leads to the following equation:

\begin{equation}
\label{eqn:ll3}
	\frac{\partial E_n}{\partial A}=\displaystyle\int\frac{\delta E_n}{\delta B(z)}\frac{\partial B(z)}{\partial A}
	\frac{dz}{1}\equiv \displaystyle\int R_n^2(z)\frac{\partial B(z)}{\partial A}dz,
\end{equation}

\noindent
where $1$ means unit length. Now, substituting Eq.~(\ref{eqn:ll3}) into Eq.~(\ref{eqn:lz}) we find analytical expression for the lateral pressure profile from the relation:

\begin{equation}
\label{eqn:lppi}
    P_t=-\int k_BT\sum_{n=0}\frac{R_n^2(z)}{E_n}\frac{\partial B(z)}{\partial A}dz\equiv \int\Pi_t(z)dz.
\end{equation} 
\noindent Hence, finally:

\begin{equation}
\label{eqn:lpp}
    \Pi_t(z)=-k_BT\frac{dB(A)}{dA}\sum_{n=0}\frac{R_n^2(z)}{E_n}.
\end{equation} 

\noindent
Calculated in our model lateral pressure profiles for the bilayer with and without interdigitation are presented in Fig.~(\ref{fig:presprof12}). It is remarkable, that lateral pressure peak at the non-interdigitated monolayers interface,
as seen in the dashed curve, disappears in the interdigitated (linked chains)case. Hence, entropic repulsion between the lipid chains is indeed weaker in the region where the entropy related with the chain orientation order is smaller (compare with Fig.~{\ref{fig:sz_12}}).

Next, it is straightforward to check that due to orthonormality of the eigenfunctions $R_n(z)$ the integral of $\Pi_t(z)$ over $dz$ across the bilayer thickness leads again to the expression in Eq.~(\ref{eqn:lz}) for the total lateral tension $P_t$:

\begin{equation}
\label{eqn:pt}
P_t=-k_BT\frac{dB(A)}{dA}\sum_{n=0}\frac{1}{E_n}\equiv-k_BT\sum_{n=0}\frac{\partial E_n}{\partial A}\frac{1}{E_n},
\end{equation}
\noindent where we used relation that follows from Eq.~(\ref{eqn:8}):

\begin{equation}
\label{eqn:eq}
\frac{dB(A)}{dA}=\frac{\partial E_n}{\partial A},\,\forall{n}.
\end{equation}

\begin{figure}[h]
	\centering
		\includegraphics{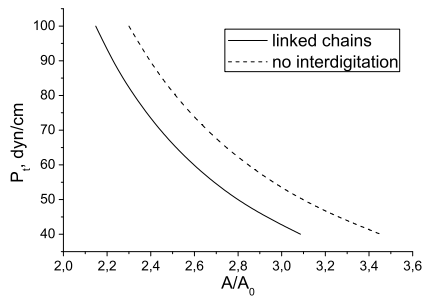}
	\caption{Total lateral pressure comparison. With the same are per chain, membrane with linked chains produce less pressure than no interdigitation membrane. }
	\label{fig:pt_12}
\end{figure}

\noindent
In Figure~(\ref{fig:pt_12}) the calculated pressure-area isotherms for interdigitated (linked chains, solid line) and non-interdigitated (dashed line) bilayer are presented. It is obvious from the figure that lateral entropic repulsion responsible for the lateral pressure in the hydrophobic part of the bilayer is weaker in the interdigitated bilayer comparatively with non-interdigitated bilayer at the one and the same area per lipid chain and other parameters fixed.

Differentiation of $P_t(A)$ gives the area compressibility modulus

\begin{equation}
\label{eqn:15}
	K_a=-A\frac{\partial P_t}{\partial A}
\end{equation}

\noindent
as a function of the area per chain and temperature. The equilibrium condition is found by equating the pressure produced by linked chains to the effective lateral pressure in the bilayer:

\begin{equation}
	\label{eqn:16}
	P_t(A(T))=P_{eff}=\gamma + P_{HG}+P_{vdW},
\end{equation}

\noindent
where $\gamma$ is the surface tension at the hydrophobic-hydrophilic interface; $P_{HG}$ is the head group repulsion of electrostatic origin; $P_{vdW}$ is the pressure arising from the vander Waals interactions between chains, etc. We choose $P_{eff}>\gamma \sim 70$ $dyn/cm$ because attractive dispersion interactions between hydrocarbon chains are included in the effective surface tension \cite{Lindahl}. At room temperature for a typical lipid bilayer with effective surface tension one has: $50\le P_{eff}\le 150$ $dyn/cm$ \cite{Marsh, Lindahl}.
Analytical solution for the total presure in case of linked chains (interdigitation):

\begin{equation}
\label{eq:1}
	P_t^{linked}= \frac{2 k_B T}{3 A_0 \nu^{1/3} \sqrt{a} (\sqrt{a}-1)^{5/3}} \cdot \left(2 \nu^{2/3}      (\sqrt{a}-1)^{2/3}  +1  \right)
\end{equation}
\noindent
Analytical solution for the total pressure in case of not linked chains (no interdigitation):

\begin{equation}
	\label{eq:2}
	P_t^{noint}= \frac{2 k_B T}{3 A_0 \nu^{1/3} \sqrt{a} (\sqrt{a}-1)^{5/3}} \cdot \left(4 \nu^{2/3}      (\sqrt{a}-1)^{2/3}  +1  \right)
\end{equation}

\noindent It follows from the analysis of these expressions that interdigitation effect on the total lateral pressure at a given area is more pronounced at larger areas per lipid (lower pressures) region, that corresponds to hihger orientational disorder of the chains.

\begin{figure}[h]
	\centering
		\includegraphics{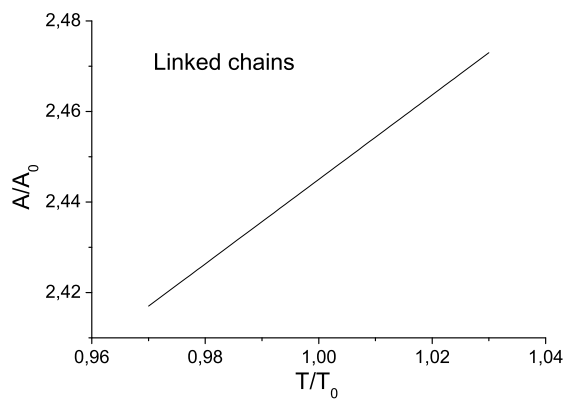}
	\caption{Tempereture dependence of equilibrium area per chain, A. Temperature is normalized by $T_0$, area is normalized by $A_0$.}
	\label{fig:aT_12}
\end{figure}

\begin{figure}[h]
	\centering
		\includegraphics{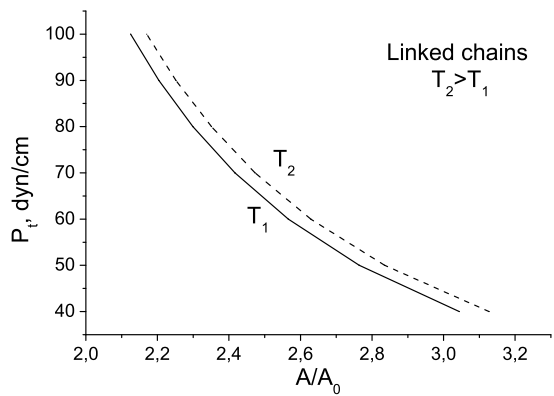}
	\caption{Calculated total lateral pressure $P_t$ produced by linked hydrocarbon chains as a function of area per chain at two temperatures $T_1$ (solid line)$<$ $T_2$ (dashed line). Lateral pressure is normalized by $k_BT/A_0$, area is normalized by $A_0$. Other parameters are as in Fig.~(\ref{fig:presprof12}). }
	\label{fig:pt_linked_12}
\end{figure}

In Fig.~{\ref{fig:aT_12}} the temperature dependence of the area per chain in the bilayer is shown. This curve is increasing with temperature due to a more frequent collisions of chains. Fig.~{\ref{fig:pt_linked_12}} displays $P_t(A)$ dependence. This curve is decreasing with the area per chain due to the following reason: when the chains occupy more space they collide less frequently and produce less entropic pressure.

Now we can verify the exploited approximation of the small chain deviations in the bilayer Eq.~(\ref{eqn:2}). We calculate the thermodynamic average of the chain fluctuation amplitude $\left\langle \textbf {R}^2(z) \right\rangle$ using the relation $\left\langle R^2_{x,y}(z)\right\rangle =\sum_n\left\langle C_n^2\right\rangle R_n^2(z)$ and averaging over $C_n$:

\begin{equation}
	\left\langle \textbf{R}^2(z)\right\rangle=k_BT\sum_n\frac{R_n^2(z)}{E_n}.
	\label{eqn:23}
\end{equation}

It is worth mentioning that integration of both sides of Eq.~(\ref{eqn:23}) over $z$ from $0$ to $2L$ provides the self-consistency equation Eq.~(\ref{eqn:13}). Since $E_n \propto n^4$, the sum in Eq.~(\ref{eqn:23}) converges fast and allowing for the relation $R_n^2(z) \sim 1/L$, we can estimate it as $\sum_n 1/E_n \propto 1/B$. According to Eq.~(\ref{eqn:et}), the increase of potential energy associated with the increase of area swept by the string from $0$ to $\delta A$ is of order $B2L\delta A$ (the string is formed by the centers of the chain cross-sections). On the other hand, it is equal to the work against the pressure $P_{eff}$ needed to increase the area per couple of linked chains in the bilayer from $A_0$ to $A$: 
$B2L\delta A \approx P_{eff}(A-A_0)$. From the last equality and relation Eq.~(\ref{eqn:a}) it follows that $B>P_{eff}/L$. Then we evaluate:

\begin{equation}
	\sum_n \frac{R_n^2(z)}{E_n} \le \frac{1}{P_{eff}},
	\label{eqn:24}
\end{equation}

and find a rough estimate for the upper limit of the small parameter:

\begin{equation}
	\label{eqn:25}
	\sqrt{\left\langle R_n^2(z)\right\rangle}/2L \le (k_BT/L^2P_{eff})^{1/2}=0.16.
\end{equation}

Finally, we compare the amplitudes of linked chains fluctuations in the bilayer Eq.~(\ref{eqn:23}) and in empty space. For a free couple of chains with flexural rigidity $K_f$ the characteristic deviation $R^0$ can be evaluated by equating the chain bending energy to $k_B T$. This yields

\begin{equation}
	\label{eqn:36}
	R^0 \propto \frac{k_BT}{K_f}L^3 \sim L^2
\end{equation}

Allowing for Eq.~(\ref{eqn:23}) and Eq.~(\ref{eqn:36}), we find $\sqrt{\left\langle R^2\right\rangle/R_0} \sim 0.1$.

\section*{Acknowledgments}
The authors acknowledge valuable discussions with Professor Yu. Chizmadzhev and co-workers at the Frumkin Institute .

\appendix

\section{Solution of self-consistency equation}

Here we present the solution of the self-consistency Eq.~(\ref{eqn:13}) and find the analytical temperature and area per couple of chains dependence of the lateral pressure produced by the linked chains using the equation of state Eq.~(\ref{eqn:14}). It is convenient to perform the derivations in dimensionless parameters,

\begin{equation}
	\label {eqna:1}
	a=A/A_0, b=\frac{L^4}{K_f}B,
\end{equation}

and to introduce the auxiliary parameters

\begin{equation}
	\label{eqna:2}
	k_n = (\pi n/2)^4, n \ge 1; v=\frac{K_f A_0}{\pi k_B T L^3},
\end{equation}

where $L \sim 15 A$ is the chain length, $A_0 \sim A^2$ is the "incompressible area" of the chain cross section, and the chain flexural rigidity $K_f\cong k_B TL/3$ at $T \approx T_0=300 K$. Using these estimates we obtain $v \cong 0.009$.

In the introduced notations Eq.~(\ref{eqna:1}) and Eq.~(\ref{eqna:2}) with $E_n$ defined in Eq.~(\ref{eqn:8}) the self-consistency equation Eq.~(\ref{eqn:13}) acquires the form

\begin{equation}
	\label{eqna:3}
	\frac{1}{b} + \sum_{n=1} \frac{1}{b+k_n} = 2v(\sqrt{a}-1)^2.
\end{equation}

The terms in the sum on the left hand side of Eq.~(\ref{eqna:3}) decrease fast with growing $n$ and we can use integration instead of summation over $n$. For example, tt the effective tension $P_{eff}=70$ $dyn/cm$, we have $b \approx 10^3$, while $k_11^4\approx 9 \cdot 10^4$. In this regime we can solve Eq.~(\ref{eqna:3}) analytically by substituting summation over $n$ with integration, which yields

\begin{equation}
	\label{eqna:4}
	\sum_{n=1}\frac{1}{b+k_n^4} \approx \frac{1}{2} \int_{-\infty}^{\infty} \frac{dn}{b+k_{n+1}^4}=\frac{1}{\sqrt{2}b^{3/4}}
\end{equation}

\noindent
where we took integral using complex functions theory, and $k_n$ is defined in Eq.~({\ref{eqna:2}}).

In case of membrane with no interdigitation (see Fig.~{\ref{fig:effect}}) Eq.~(\ref{eqna:3}) takes the form:

\begin{equation}
	\label{eqna:5}
	\frac{1}{b} + \sum_{n=1} \frac{1}{b+k_n^4} = v(\sqrt{a}-1)^2.
\end{equation}

\noindent
Since $b \approx 10^3$ ($P_{eff}=70$ $dyn/cm$), we (approximately) integrate over $n$, so that in case of membrane with no interdigitation Eq.~(\ref{eqna:4}) takes the form:

\begin{equation}
	\label{eqna:6}
	\sum_{n=1} \frac{1}{b+k_n^4} \approx \frac{1}{2} \int_{-\infty}^{\infty} \frac{dn}{b+k_{n+1}^4}=\frac{1}{2\sqrt{2}b^{3/4}}.
\end{equation}

\noindent
where $k_n$ is defined as in \cite{mubao}: $k_n=\pi n-\pi/4$.

In both, Eq.~(\ref{eqna:4}) and Eq.~(\ref{eqna:6}), we omit $1/b$ as $b \approx 10^3$, which leads to the same $b(a)$ dependences in both cases:

\begin{equation}
	\label{eqna:7}
	b= \frac{1}{4v^{4/3}(\sqrt{a} -1)^{8/3}},
\end{equation}

which is then used in the equation of state Eq.~(\ref{eqn:14}). As a result we find the expression for the lateral pressure produced by the linked hydrocarbon chains Eq.~(\ref{eqn:14})

\section{Trans-membrane structure of rafts}

\begin{figure}[h]
	\centering
		\includegraphics{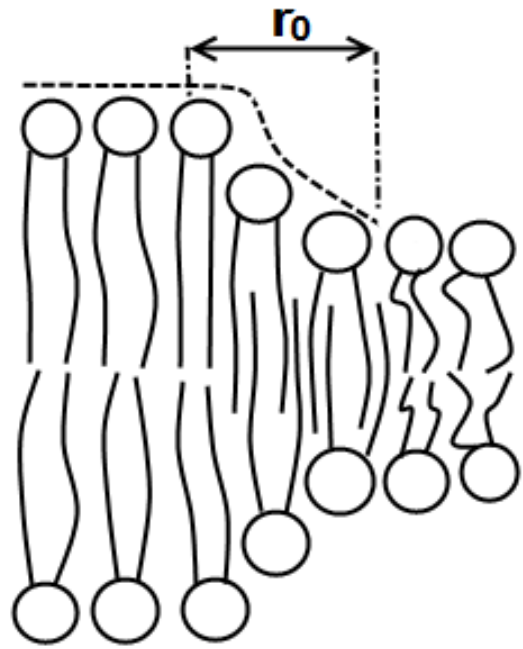}
	\caption{Assumed raft boundary structure.}
	\label{fig:prof}
\end{figure}

Here we make some estimates to find out if there could be free energy gain related with interdigitation along the perimeter of hydrophobically mismatched regions in the membrane, e.g. for the case of a cluster of the ordered lipids surrounded by the liquid membrane ``sea'' (model of a raft). We evaluate energy of hydrophobic mismatch using the following expression:

\begin{equation}
	\label{eqn:28}
   F_{LP}=\frac{\pi}{4} (K_d r_0^2 + K_g)(\frac{2r_p}{r_0}+1)(d_p-L)^2
\end{equation}

where $r_p \sim 25 A$ \cite{Sukharev} is a protein radius, $r_0 \sim 10 A$ \cite{Ben-Shaul} is a characteristic scale of deformation (see Fig.~\ref{fig:prof}), $K_d \sim 15 \cdot 10^14$ $erg/cm^4$ \cite{Hamill} is a dilation modulus, 
$K_g \sim 35$ $erg/cm^2$ \cite{Fournier} is the modulus, characterizing the energy cost of producing a gradient of bilayer thickness (it includes the energy of increasing the area of chain-water interface), $L$ is the (equilibrium) monolayer thickness, $d_p$ is the protein monolayer thickness. One can see $F_{LP}$ values with given values of parameters and various $d_p - L$, i.e. raft-bilayer thickness micmatch.

\begin{table}
	\centering
		\begin{tabular}{ c|c }
			\hline
			\hline
			$d_p-L$, A  &  $F_{LP}$, $k_B T$ \\
			\hline
			{1} &  {0.61} \\
			{2} &  {2.43} \\
			{3} &  {5.46} \\
			{4} &  {9.71} \\
			{5} &  {15.17} \\
			{6} &  {21.84} \\
			\hline
		\end{tabular}
	\caption{Energy of hydrophobic mismatch ($F_{LP}$ see Eq.~\ref{eqn:28}) as a function of raft thichness (L is a lipid membrane monolayer thickness, i.e. constant)}
	\label{tab:1}
\end{table}

Thus, we found that $F_{LP} \sim 10 k_B T$ whereas energy cost of interdigitation in the $r_0$ area (see Fig.~\ref{fig:prof}) is $F_{int} \sim 100 k_B T$ (one can easily obtain this value by counting chain quantity in the $r_0$ area and multiply its by $\Delta F_{int}^{per chain}$). This means, that, within our model, interdigitation itself does not help to decrease the hydrophobic mismatch energy of the cluster (raft).

%\bibliography{smkb1}

% Figure legends
\clearpage
\section*{Figures}

\end{document}